# Evaluation and insights from a sonification-based planetarium show intended for improving inclusivity

Chris Harrison (Newcastle University)*, Anita Zanella (INAF) and Aishwarya Girdhar (Newcastle University)
*Contact: christopher.harrison@newcastle.ac.uk

## Abstract

Audio Universe: Tour of the Solar System is an audio-visual show for planetariums and flatscreen viewing. It is designed in collaboration with members of the blind and vision impaired (BVI) community, BVI specialist teachers and their pupils. It aims to be suitable for audiences with all sight levels by representing key concepts through sound and using a carefully constructed narration. We present results from 291 audience evaluations from online viewers and audience members of several planetarium showings in the UK and Italy. We find a strong appreciation from BVI and non-BVI audiences, with ~90% scoring 4 or 5 (out of 5) for both how useful and enjoyable the sounds are. We also present results from surveying planetariums and communication leaders known to have downloaded the show. We find international success for special events, for BVI audiences and for those with other special educational needs and disabilities (SEND; including sensory needs and learning difficulties). Feedback suggests this is due to its multi-sensory, clearly narrated, and low sensory load (calm) production. However, we also describe limitations identified during this evaluation exercise, including the show's limited incorporation into regular (non-special) planetarium programmes. This highlights an ongoing challenge of creating a fully inclusive planetarium experience.

## 1. Introduction

Astronomy communication is blessed with inspiring images. This is highlighted by the recent flurry of images from the *James Webb Space Telescope*, which have reached top media forums internationally. Furthermore, computer simulations have produced animations and images popular for public communications, including visualisations of "invisible" phenomena, such as dark matter. This all lends itself to producing visually captivating planetarium shows, combining observational data, simulated data and artistic impressions. However, this visually-focused approach naturally excludes audiences who are blind or vision impaired (BVI) and, more generally, those who prefer non-visual communication.

Innovative approaches are needed to make astronomy communication and research more BVI accessible (e.g., Pérez-Montero, 2019; Noel-Storr & Willebrands, 2022; Foran, Cooke & Hannam, 2022). Multi-sensory approaches to astronomy communication open a route to be more inclusive and to make the content more engaging for everyone. Encouragingly, in recent years, there has been an increase in astronomy communication projects which focus on tactile resources (e.g., Bonne et al., 2018; Paredes-Sabando & Fuentes-Muñoz, 2021; Arcand & Watzke, 2022) and/or audible resources (e.g., Quinton et al. 2016; Tomlinson et al., 2017; Bieryla et al., 2020; Elmquist et al., 2021; García-Benito & Pérez-Montero, 2022; Bardelli et al., 2022). Turning astronomical data into sound, a process called "sonification", has seen a particular boost in popularity over recent years for applications in astronomy research and communication (see review in Zanella et al., 2022).

Whilst encouraging, there are some limitations in preventing wider adoption of the developed multi-sensory and BVI-accessible astronomy communication resources and methods. For example: (1) many have been developed for a small number of one-time events run by the developers, without the resources widely shared; (2) they require tactile or other specific physical resources which can be difficult to manufacture or are too time-consuming or costly to produce for large audiences; and (3) they require a live presenter who would need to be trained and comfortable in delivering the specialist resources. Therefore, we identified a need for a pre-rendered BVI-accessible astronomy show that is easy to disseminate internationally and requires little-to-no effort for planetarium presenters or other astronomy communicators to use. To this aim, we created, as part of our broader Audio Universe project [*1], the BVI-accessible show, *Audio Universe: Tour of the Solar System*. In this article, we discuss the evaluation results and lessons learnt since the show's launch in December 2021.

## 2. Design and creation of the show

The design process for *Audio Universe: Tour of the Solar System* and a description of the methods used are reported in Harrison et al. (2022). However, in this section, we briefly summarise the design process and methods used to help put our new evaluation results into context.

Our target audience was primary school children (aged 7-11 years); therefore, we included educational content about the Solar System. Our goal was to create a show that planetariums and wider astronomy communicators could use to attract school groups whilst also appealing to general family audiences, irrespective of the level of vision of the visitors (from fully sighted to fully blind). We produced a full-dome and flat-screen version, a surround sound, and a stereo soundtrack.

Our design focus was on the soundtrack, which was to be understandable without the associated visuals. During design and development, we worked collaboratively with focus groups, including members of the BVI community and teachers of BVI pupils and young people, as well as the BVI pupils themselves. We also worked with a music composer and a creative writer (see Harrison et al., 2022).

The theme for the show is a tour aboard a special spacecraft, which is fitted with a "sonification machine". This machine turns light into sound, and at each location on the journey around the Solar System, the objects are represented with sounds. For example, the planets are represented by different musical instruments, with pitches chosen to correspond to their mass. In a show segment, the surround sound helps create the impression that the planets can be heard orbiting around the audience. The sounds were produced using actual data and the STRAUSS code [*2], also used in astronomy research (e.g., Tucker-Brown et al., 2022; Trayford et al., 2023a,b). The work with the focus groups was crucial. For example, we learnt the importance of slow-paced and extremely descriptive narration (but avoiding too much reliance on visual metaphors) and giving advance warning of what was coming next. In contrast to traditional show production, we completed the soundtrack first. The visuals were added later by a professional planetarium producer.

*Audio Universe: Tour of the Solar System* was released in December 2021 in English, Italian and Spanish (a version in German is now available). The narration is performed by the "spaceship's captain" and an expert blind astronomer. The latter is real-life Australian-French blind astronomer Dr Nic Bonne in the English version. A voice actor plays this character in the Italian and German versions. In the Spanish version, the real-life Spanish blind Astronomer Dr

Enrique Pérez-Montero is used instead (played by a voice actor). These characters provide role models for BVI people watching the show.

## 3. Dissemination, reach and evaluation approach

A series of premier events across the UK and Italy were planned to coincide with the International Day of Disabled Persons in December 2021. Simultaneously, we released the show for free download, in both full-dome and flatscreen format, and uploaded it to YouTube[*3].

### *3.1 Dissemination and usage information*

Downloading the show for anything other than personal use comes with the request to complete a Google Form to collate basic information about the proposed use, contact details, and an agreement to record information about the show's use on a best-effort basis. Although this is stated as compulsory on the download page [*3], it was impossible to require the completion of the form to initiate the download. Indeed, internet searches have revealed that the show has been used by institutes that did not complete the form.

As of 15 February 2023, 51 institutions completed the form, and a further eight institutions were identified as having used the show due to email contacts or internet searches by the authors. These institutions cover 16 countries, including 46 fixed planetariums, 11 travelling science communicators (mostly those using portable planetariums), one cinema, and one school. We followed up with these 59 institutes by sending emails from 15 - 28 February 2023. We asked each of these institutes how they had used the show or planned to use it in the future. We received responses from 35. We discuss the qualitative feedback received in Section 4.2.

Based on combined YouTube viewing statistics obtained on 24 April 2023, the show had 4987 views across the various channels it has been uploaded [*3].

### *3.2 Audience Evaluations*

We produced audience feedback forms on paper and online in both English and Italian. We focussed our evaluation efforts on the UK and Italy because these are the countries of the team running these evaluations (i.e. the authors of this article). However, our YouTube channel also advertised an online form for completion [*3].

The evaluation form questions are provided in full in Info Box 1, including demographic and feedback questions. The feedback questions focus on the perception of BVI accessibility, whether or not the audience enjoyed the show and their opinions of how useful the sound representations were. In Section 4.1, we present the results of this evaluation.

## 4. Results and Discussion

Here, we discuss the feedback received between December 2021 and February 2023 from people who watched the show (Section 4.1) and institutes that downloaded the show (Section 4.2).

### *4.1 Audience Feedback*

We received 291 completed audience evaluation forms. Live audience members were asked to complete paper forms after watching the show. These were handed out by those running the events. BVI audience members who were unable to complete the form themselves were directed to the online version of the form, or their answers were transcribed by an assistant. Online viewers could complete the form if they followed a link in the YouTube video description.

The majority (253) of those who completed the forms saw the show in planetariums or other venues (204 in Italy, 47 in the UK, and two in the USA), and the remainder (38) viewed the show online. A summary of the demographics of those who completed the survey is presented in Table 1. We particularly note the high fraction (relative to the general population) of those who self-declared as BVI (25%) and those involved in the care, education or welfare of BVI individuals (22%). For example, from a study in 2015, 0.49% and 2.95% of the world population are blind or have moderate to severe visual impairment, respectively (Bourne et al., 2017; Ackland et al., 2017). Another important part of the demographics is that the majority (78%) are older than 18. In a future publication, we will provide the results of a separate, dedicated evaluation exercise of school children who experienced the show in addition to other multisensory activities in educational settings.

We found a very high overall rating from audience members of *Audio Universe: Tour of the Solar System*, with an average score of 4.48+/-0.04 out of 5 (uncertainty is the standard error on the mean), with 91% giving a score of 4 or 5. The overall positive perception of the sound-based aspects of the show is also clear, with 90% scoring a 4 or 5 for how enjoyable they found the sounds (average score 4.50+/-0.05) and 87% scoring a 4 or 5 for how *useful* they found the sounds (average score 4.45+/-0.05). Furthermore, 74% responded that they were now more convinced that astronomy is accessible to people who are BVI after watching the show, and 82% responded "Yes" to wanting to find out more about science after watching the show.

In Table 2, we report the results for questions 2, 3 and 4 of the feedback questions across different demographic groups. These are the most interesting questions for the present study: to understand the audience's opinions on the novel use of sound in this show and their perception of BVI accessibility. Overall, we found little variation in the feedback responses from the different demographic groups. We take this as a positive outcome: this show was perceived as enjoyable and interesting and promoted accessibility for all audiences. However, we highlight some particular results in Figures 1 and 2 and the following text.

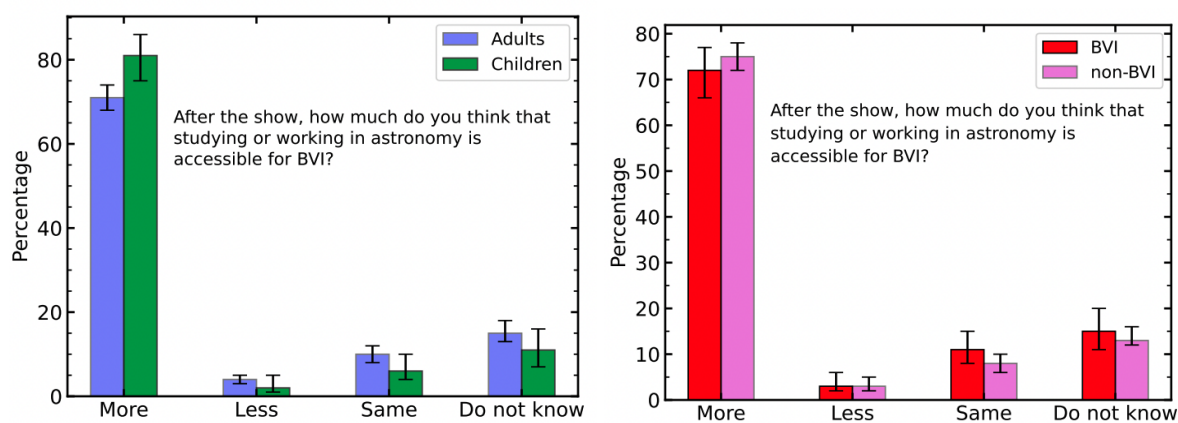

**Figure 1**: Distribution of answers to the question, After watching Audio Universe: Tour of the Solar System, how much do you think that studying or working in astronomy is accessible for people who are blind or vision impaired? *Left panel*: answers of adults (blue histogram) compared to children (< 18 years old, green histogram). *Right panel*: answers of BVI (red filled histogram) compared to non-BVI (magenta histogram) audience. The error bars indicate 68.3% Wilson score binomial confidence intervals.

Figure 1 (left panel) shows that children are more likely than adults to be more convinced that astronomy is accessible for people who are BVI after watching the show: specifically, $81^{+5}_{-6}$% for children compared to $71^{+3}_{-3}$% for adults (where ranges indicate 68.3% Wilson binomial confidence intervals). It is also worth highlighting that $76^{+5}_{-6}$% of audience members involved in the education/welfare of BVI people were more convinced about BVI accessibility in astronomy after the show. It is a positive outcome that such a high fraction of those involved in caring for BVI people are now more positive about accessibility in astronomy after the show.

Figure 1 (right panel) shows that BVI and non-BVI audiences show a similar fraction of answering positively to this same question about accessibility ($72^{+5}_{-6}$% and $75^{+3}_{-3}$%, respectively). We take this as a positive sign that this show and similar efforts can have a positive impact on perceptions of accessibility for all audience members. Indeed, across most questions, the BVI and non-BVI audience members scored very similarly (Table 2). However, for the question "How useful did you find it to have the objects in space represented by sounds to understand the show?", there is a marginal difference, highlighted in Figure 2. The "very useful" option was selected by $73^{+5}_{-5}$% of the BVI audience and $61^{+3}_{-3}$% of the non-BVI audience. It may be unsurprising that the sounds are relatively more useful for the BVI audiences; however, it is interesting to note that significantly more than half of the non-BVI audience also answered "very useful" to this question. We believe that this emphasises the benefits of creating multi-sensory, accessible shows for all audiences and that considering such endeavours should not just be limited to BVI audiences.

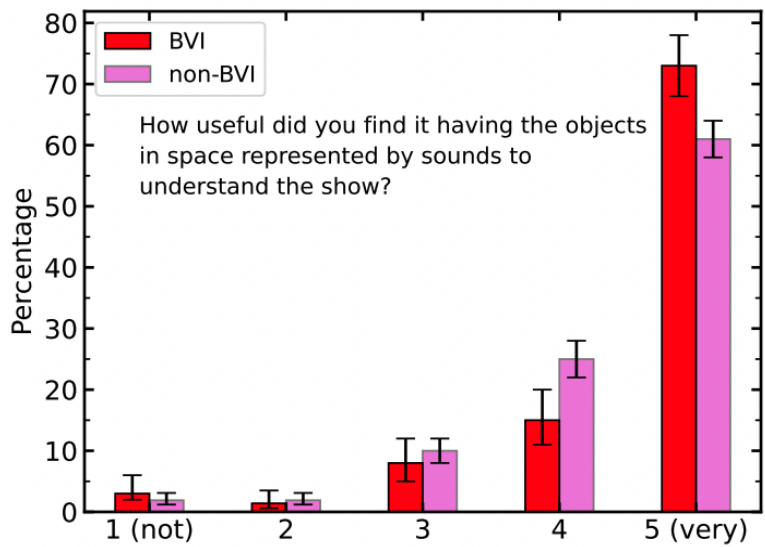

**Figure 2**: Distribution of answers to the question: How useful did you find it to have the objects in space represented by sounds to understand the show? A comparison of the answers from BVI (red histogram) and non-BVI (magenta histogram) audience members. The error bars indicate 68.3% Wilson score binomial confidence intervals.

The final part of the evaluation form asked for "any other comments". "Sound" was the most mentioned word in this feedback, with many positive descriptive words to describe the sounds and show overall, including "beautiful", "interesting", and "amazing". Most of the more critical comments were about sound balance ("too loud", "not enough bass", etc.). However, as the specifics of these issues were inconsistent, they were likely related to the specific venues in which they heard the show. Nonetheless, this highlights how important the balance in the soundtrack is and that guidance could be provided to support venues for fine-tuning their sound system for sound-based shows. Another comment, which appeared a few times, was the perceived difficulty in audibly distinguishing the different planet sounds when they were heard together. Although this did not come up during the focus group testing phase (Harrison et al., 2022), further experimentation with different sounds and wider testing could help assess the audience's ability to distinguish between multiple sounds heard simultaneously in future shows.

The importance of the soundtrack and the show's overall accessibility to BVI audience members is exemplified by the qualitative feedback from members or carers of the BVI groups. One online viewer who lost their sight as a child said:
> *"Had the best and greatest accessible experience of my entire life. The sounds gave me goosebumps".*

A live BVI audience member in Italy said:
> *"I am blind from birth and however much I have always tried to imagine how the space that surrounds us was made, only now thanks to you, I was able to get a clear idea of what is around us ... and it's beautiful!".*

A live BVI audience member in the UK said:
> *"Before I went I thought planets, stars and space were all for sighted people but when I went there [the show] proved me wrong. And I then thought, oh wow anything is accessible and it was really good. Because usually when you go to a planetarium or something like that it is often very visual and I'm like 'what is going on?'".*

## *4.2 Planetarium and Communication Leaders' Feedback*

Of the 35 institutions (predominantly planetariums) which responded to our enquiries, 26 have already used the show, and nine have not yet presented the show. However, most of these nine anticipated using the show in the future for specifically planned events or when a version of the show can be produced in the local language. Some of those who did not use the show discussed concerns with the directional language (pertaining to the use of "left" and "right") because their planetariums have circular seating, which can lead to the audiences having different perspectives. This has been noted as something that can be improved or avoided in future productions. One other planetarium stated that they did not intend to use the show because they were not sure how to market the show's title and content for their planetarium.

There are common themes in how the show is being used internationally. The show is predominantly used for specific events for BVI and other SEND audiences. The feedback has been extremely positive for being able to use the show for these groups, with gratitude for providing this resource for free. Although the original primary focus was on BVI audiences, the anecdotal feedback (including some comments in the evaluation forms) we have received highlights that other SEND audiences have also benefited. This includes those with sensory needs and learning difficulties. This appears to be due to the multi-sensory, calm nature and clear narration of the show. For example, one planetarium leader said:

> *"I have used it on three separate occasions already. Two of them were visits to SEND schools, where I discovered, through trial and error, that the show is calm enough not to cause sensory overload, while also being multi-sensory (sight and sound matching up well), which is something SEND schools in particular appreciate a lot. The last time I used the show…was for two SEND classes within a traditional school: I presented the show to the teachers as one of the options in our repertoire, and the calm aspect of it appealed to them greatly."*

Despite the positive feedback and high uptake of the show for specially planned BVI and other SEND events, we are only aware of two planetariums that have incorporated this show into their regular programs of public events for all audiences (e.g., a daily, weekly or monthly regular planetarium program). The reasons provided for this include more minor and easily addressable comments (for future shows); for example, the show is more than 30 minutes, which is not suitable for the timetabling approach in some planetariums. More broadly, a few planetarium directors felt the show was not suitable for their usual family audiences because of its slow pace compared to other family shows. There also appears to be some perception that this show is exclusively suitable for BVI and SEND audiences. As far as we know, this is based on the opinion of the planetarium leaders rather than directly surveying their own audiences. At face value, this appears to be in contrast with our audience evaluations (Section 4.1), for which there was a high level of enjoyment of the show and appreciation of the sounds across all demographic groups.

As of this writing, we have not achieved one of our objectives, as most planetariums do not schedule this show into their regular programming. We aimed to create a show that was considered fully inclusive and suitable for all audiences (irrespective of level of vision). Whereas, with only a couple of exceptions, the current status is that it is being used (albeit successfully) exclusively for BVI and SEND audiences.

## 5. Conclusions

We have presented our evaluation of the show *Audio Universe: Tour of the Solar System*. Overall, these results suggest that all audiences, irrespective of their level of vision, found the novel sound-based approach useful and enjoyable. Furthermore, it has had a positive impact on their perception of BVI-accessibility in astronomy. We have had the unanticipated outcome that the show has benefited wider SEND audiences. Therefore, we feel encouraged that further shows taking a similar approach to *Audio Universe: Tour of the Solar System* will benefit inclusivity and accessibility in planetariums and other astronomy communication settings. The qualitative feedback we have received will help fine-tune future shows.

We note that our results are biased towards Italian and UK audiences, and other studies would be required to understand the best approaches to using sounds elsewhere. Importantly, we cannot necessarily extrapolate these conclusions to the perceptions of audiences with particularly different musical cultures due to the predominantly Western musical choices made during this show's creation (e.g., McDermott et al., 2016, García-Benito & Pérez-Montero, 2022).

Our results from surveying planetarium directors are mostly encouraging, with an international reach and positive outcomes for special events for BVI and other SEND groups. Future work may break down the analysis further into different types, or levels, of vision impairment and other specific subsets of educational needs and disabilities of the audience members.

We found that most planetariums are not using this show for their general programmes, which means that the show is not being used for fully inclusive events. Future accessible show development should consider how to overcome this limitation by working directly with planetarium directors and their audiences. This may involve some changes to the approach of the show's content, design and style. However, it may also involve changes to how the show is marketed, including how the benefits and enjoyment for all audiences are demonstrated to planetarium directors.

We now encourage show developers to explore using our STRAUSS code[*2] to develop their own sonifications. The code is released on GitHub with some example scripts used to create the sonifications for *Audio Universe: Tour of the Solar System*.

**Notes**

*1 Audio Universe website: [www.audiouniverse.org](www.audiouniverse.org)

*2 STRAUSS code: https://github.com/james-trayford/strauss

*3 Locations the show can be downloaded or viewed online: (1) European Southern Observatory's (ESO's) archive full-dome version (https://www.eso.org/public/videos/au-totss-fulldome/): (2) ESO's archive flat-screen version (https://www.eso.org/public/videos/au-totss/); (3) Audio Universe project's YouTube channel (https://www.youtube.com/@audiouniverse8137).

**Acknowledgements**

We acknowledge funding from a United Kingdom Research and Innovation grant (MR/V022830/1), a Science and Technology Facilities Council Spark Award (ST/V002082/1) and a Royal Astronomical Society Education and Outreach Small Grant. We thank all of the audience members and planetarium or communication leaders who took the time to complete feedback forms and respond to our enquiries.

**Table 1**

Demographic summary of the audience members who filled in the evaluation forms.

| Self-identify as | Per cent | Age | Per cent | Gender | Per cent |
|---|---|---|---|---|---|
| BVI | 25% | < 7 | 2% | Female | 56% |
| BVI parent/carer or involved in BVI education/welfare (combined) | 22% | 7-10 | 7% | Male | 42% |
| STEM-related job or involved in science education/communication (combined) | 25% | 11-14 | 7% | Non Binary | 0.3% |
| Teacher | 14% | 15-18 | 3% | Prefer not to say | 1.4% |
|  |  | >18 | 78% |  |  |
|  |  | Prefer not to say | 3% |  |  |

## Info Box 1: Audience feedback form questions in full.

Part 1: Demographic Questions.
- (1) Where did you see the show?
- (2) What is your age bracket?
  [options: <7, 7-10,11-14,15-18, >18, Prefer not to say]
- (3) What is your gender
  [options: Female, Male, Non binary, Prefer not to stay]
- (4) Do you identify as someone who is blind or vision impaired?
  [options: Yes, No, Prefer not to say]
- (5) Do you identify as any of the following?
  a. Parent/carer of somebody who is blind or vision impaired
  b. Involved in the education or welfare of people who are blind or vision impaired (except parent/carer)
  c. Teacher
  d. Working/worked in a job related to science, technology, engineering or mathematics (STEM)
  e. Involved in science communication/education (other than teachers)

Part 2: Feedback Questions.
- (1) After watching Audio Universe: Tour of the Solar System, do you want to find out more about science?
  [options: Yes, No, Don't know]
- (2) After watching Audio Universe: Tour of the Solar System, how much do you think that studying or working in astronomy is accessible for people who are blind or vision impaired?
  [options: "I am now more sure that astronomy is accessible for people who are blind or vision impaired"; "I am now less sure that astronomy is accessible for people who are blind or vision impaired"; "My opinion is unchanged"; "Don't know"]
- (3) On a scale of 1 to 5, how much did you enjoy having the objects in space represented by sound? [1 = did not enjoy; 5 = very much enjoyed]
- (4) On a scale of 1 to 5, how useful did you find it to have the objects in space represented by sounds to understand the show? [1 = not at all useful; 5 = very useful]
- (5) On a scale of 1 to 5, overall, how would you rate Audio Universe: Tour of the Solar System? [1 = poor; 5 = excellent]
- (6) Please provide any other comments here.

**Table 2**
Results of the audience feedback questions 2, 3 and 4 split into various demographic groups. Numbers are percentages, and the upper/lower bounds indicate 68.3% Wilson score binomial confidence intervals. The final category, "Other", includes all people who did not identify themselves in any category of question (5) of the demographic questions (see Info Box 1).

| | Adults | Children | BVI | Non-BVI | BVI carers/welfare | Science related job/coms. | Teachers | Other |
|---|---|---|---|---|---|---|---|---|
| **After Audio Universe: Tour of the Solar System, how much do you think that studying or working in astronomy is accessible for people who are blind or vision impaired?** | | | | | | | | |
| More | $71^{+3}_{-3}$ | $81^{+5}_{-6}$ | $72^{+5}_{-6}$ | $75^{+3}_{-3}$ | $76^{+5}_{-6}$ | $74^{+5}_{-5}$ | $66^{+7}_{-8}$ | $75^{+3}_{-4}$ |
| Less | $4^{+1}_{-1}$ | $2^{+3}_{-1}$ | $3^{+3}_{-1}$ | $3^{+1}_{-1}$ | $0^{+2}_{-0}$ | $5^{+3}_{-2}$ | $0^{+2}_{-0}$ | $4^{+2}_{-1}$ |
| Same | $10^{+2}_{-2}$ | $6^{+4}_{-2}$ | $11^{+4}_{-3}$ | $8^{+2}_{-2}$ | $11^{+5}_{-3}$ | $7^{+4}_{-2}$ | $20^{+7}_{-5}$ | $8^{+3}_{-2}$ |
| Don't know | $15^{+3}_{-2}$ | $11^{+5}_{-4}$ | $15^{+5}_{-4}$ | $13^{+3}_{-2}$ | $13^{+5}_{-4}$ | $14^{+4}_{-3}$ | $15^{+6}_{-5}$ | $13^{+3}_{-3}$ |
| **How much did you enjoy having the objects in space represented by sound?** | | | | | | | | |
| 1 | $2^{+1}_{-1}$ | $2^{+3}_{-1}$ | $3^{+3}_{-1}$ | $1.4^{+1.1}_{-0.6}$ | $0^{+2}_{-0}$ | $0^{+1}_{-0}$ | $2^{+2}_{-4}$ | $3^{+1}_{-2}$ |
| 2 | $0.9^{+0.9}_{-0.4}$ | $0^{+2}_{-0}$ | $0^{+1}_{-0}$ | $1.4^{+1.1}_{-0.6}$ | $0^{+1}_{-0}$ | $0^{+1}_{-0}$ | $0^{+2}_{-0}$ | $2^{+2}_{-1}$ |
| 3 | $8^{+2}_{-2}$ | $6^{+4}_{-2}$ | $8^{+4}_{-3}$ | $7^{+2}_{-2}$ | $10^{+4}_{-3}$ | $7^{+4}_{-2}$ | $10^{+6}_{-4}$ | $6^{+2}_{-2}$ |
| 4 | $24^{+3}_{-3}$ | $33^{+7}_{-6}$ | $26^{+5}_{-5}$ | $25^{+3}_{-3}$ | $29^{+6}_{-5}$ | $23^{+5}_{-5}$ | $20^{+5}_{-7}$ | $28^{+4}_{-4}$ |
| 5 | $66^{+3}_{-3}$ | $59^{+6}_{-7}$ | $64^{+5}_{-6}$ | $65^{+3}_{-3}$ | $62^{+6}_{-6}$ | $70^{+5}_{-6}$ | $68^{+7}_{-8}$ | $61^{+4}_{-4}$ |
| **How useful did you find it having the objects in space represented by sounds to understand the show?** | | | | | | | | |
| 1 | $3^{+1}_{-1}$ | $0^{+2}_{-0}$ | $3^{+3}_{-1}$ | $1.9^{+1.2}_{-0.7}$ | $2^{+2}_{-1}$ | $1^{+2}_{-1}$ | $5^{+5}_{-2}$ | $2^{+2}_{-1}$ |
| 2 | $1.3^{+1.0}_{-0.6}$ | $2^{+3}_{-1}$ | $1.4^{+2.1}_{-0.8}$ | $1.9^{+1.2}_{-0.7}$ | $0^{+2}_{-0}$ | $1.4^{+2.1}_{-0.8}$ | $0^{+2}_{-0}$ | $3^{+2}_{-1}$ |
| 3 | $10^{+2}_{-2}$ | $9^{+5}_{-3}$ | $8^{+4}_{-3}$ | $10^{+2}_{-2}$ | $8^{+4}_{-3}$ | $11^{+4}_{-3}$ | $5^{+5}_{-2}$ | $9^{+3}_{-2}$ |
| 4 | $23^{+3}_{-3}$ | $24^{+6}_{-5}$ | $15^{+5}_{-4}$ | $25^{+3}_{-3}$ | $29^{+6}_{-5}$ | $24^{+5}_{-5}$ | $20^{+7}_{-5}$ | $22^{+4}_{-3}$ |
| 5 | $63^{+3}_{-3}$ | $65^{+6}_{-7}$ | $73^{+5}_{-5}$ | $61^{+3}_{-3}$ | $62^{+6}_{-6}$ | $62^{+5}_{-6}$ | $71^{+7}_{-8}$ | $63^{+4}_{-4}$ |